\documentclass[journal,twoside,web]{ieeecolor}
\usepackage{generic}
\usepackage{cite}
\usepackage{amsmath,amssymb,amsfonts}
\usepackage{algorithmic}
\usepackage{graphicx}
\usepackage{algorithm,algorithmic}
\usepackage{hyperref}
\hypersetup{hidelinks=true}
\usepackage{textcomp}
\usepackage{multirow}   
\usepackage{array}      
\usepackage{booktabs}   
\usepackage{arydshln}  

\def\BibTeX{{\rm B\kern-.05em{\sc i\kern-.025em b}\kern-.08em
    T\kern-.1667em\lower.7ex\hbox{E}\kern-.125emX}}
\begin{document}
\markboth{}{}      
\title{Unsupervised patch-based dynamic MRI reconstruction using learnable tensor function with implicit neural representation}
\author{Yuanyuan Liu, Yuanbiao Yang, Jing Cheng, Zhuo-Xu Cui, Qingyong Zhu, Congcong Liu, Yuliang Zhu, Jingran Xu, Hairong Zheng \IEEEmembership{Senior Member, IEEE}, Dong Liang, \IEEEmembership{Senior Member, IEEE},   Yanjie Zhu, \IEEEmembership{Senior Member, IEEE}
\thanks{This study was supported in part by the National Key R\&D Program of China nos. 2021YFF0501402, National Natural Science Foundation of China under grant nos.  62322119, 62201561, 62531024, 62331028, 82430062, 12226008, and 62125111, the Guangdong Basic and Applied Basic Research Foundation under grant nos. 2025A1515012966 and 2023A1515110476 and the Shenzhen Science and Technology Program under grant no. JCYJ20220818101205012.(Corresponding author: Dong Liang, Yanjie Zhu)}
\thanks{This work involved human subjects. All in-house studies were approved by the Ethics Committee of the Shenzhen Institutes of Advanced Technology, Chinese Academy of Sciences (Application No. SIAT-IRB-200315-H0455), and conducted in accordance with the Declaration of Helsinki. The publicly available OCMR dataset had obtained ethical approval in its original study.}
\thanks{Yuanyuan Liu and Yuanbiao Yang contributed equally to this manuscript.}
\thanks{Yuanyuan Liu, Yuanbiao Yang, Jing Cheng, Jingran Xu, and Yanjie Zhu are with the Paul C. Lauterbur Research Center for Biomedical Imaging, Shenzhen Institutes of Advanced Technology, Chinese Academy of Sciences, Shenzhen, China (e-mail:\{liuyy, yb.yang, jing.cheng, jr.xu, yj.zhu\}@siat.ac.cn).}
\thanks{Zhuo-Xu Cui, Qingyong Zhu, Congcong Liu, Yuliang Zhu, and Dong Liang are with the Research Center for Medical AI, Shenzhen Institutes of Advanced Technology, Chinese Academy of Sciences, Shenzhen, China (e-mail:\{zx.cui, qy.zhu, cc.liu, yl.zhu, dong.liang\}@siat.ac.cn).}
\thanks{Hairong Zheng is with Nanjing University, Nanjing, Jiangsu, China (e-mail: Hr.zheng@nju.edu.cn
).}}

\maketitle

\begin{abstract}
Dynamic MRI suffers from limited spatiotemporal resolution due to long acquisition times. Undersampling $k$-space accelerates imaging but makes accurate reconstruction challenging. Supervised deep learning methods achieve impressive results but rely on large fully sampled datasets, which are difficult to obtain. Recently, implicit neural representations (INR) have emerged as a powerful unsupervised paradigm that reconstructs images from a single undersampled dataset without external training data. However, existing INR-based methods still face challenges when applied to highly undersampled dynamic MRI, mainly due to their inefficient representation capacity and high computational cost. To address these issues, we propose TenF-INR, a novel unsupervised framework that integrates low-rank tensor modeling with INR, where each factor matrix in the tensor decomposition is modeled as a learnable factor function. Specifically, we employ INR to model learnable tensor functions within a low-rank decomposition, reducing the parameter space and computational burden. A patch-based nonlocal tensor modeling strategy further exploits temporal correlations and inter-patch similarities, enhancing the recovery of fine spatiotemporal details. Experiments on dynamic cardiac and abdominal datasets demonstrate that TenF-INR achieves up to 21-fold acceleration, outperforming both supervised and unsupervised state-of-the-art methods in image quality, temporal fidelity, and quantitative accuracy.
\end{abstract}


\section{Introduction}
\label{sec:introduction}
\IEEEPARstart{D}{YNAMIC} magnetic resonance imaging (MRI) plays an important role in clinical applications by capturing both spatial structures and dynamic cardiac motion. However, achieving high spatiotemporal resolution in dynamic MRI remains challenging due to the limited scan time. Therefore, accelerating dynamic MRI through $k$-space undersampling has attracted considerable interest.

To reconstruct images from undersampled $k$-space data, compressed sensing (CS) has been employed by exploiting sparsity and low-rank priors in dynamic image series \cite{conventional_CS}. Early approaches leveraged sparsity priors using sparsifying transforms, such as the temporal Fourier transform and spatiotemporal total variation (TV) \cite{feng2016xd,jung2009k,caballero2014dictionary}. Subsequently, the low-rank nature of spatiotemporal correlations in dynamic MRI was utilized to enhance reconstruction performance. A prevalent strategy involves arranging temporal frames into a Casorati matrix or higher-order tensor, and reconstructing images by enforcing low-rankness through matrix completion or tensor decomposition \cite{L+S,lingala2011accelerated,christodoulou2018magnetic,zhao2012image}. To further enhance low-rankness, patch-based reconstructions exploit non-local spatial redundancies by grouping similar image or volume patches into tensors, yielding sparser representations than those derived from entire image volumes, thereby achieving improved reconstruction quality \cite{yoon2014motion,phair2023free,liu2023accelerating}.

In the past decade, deep learning (DL) techniques have demonstrated substantial potential for accelerating dynamic MRI, achieving superior reconstruction performance compared with CS-based reconstructions. Generally, two main categories of DL-based methods have been explored: supervised and unsupervised learning methods. Supervised methods utilize extensive fully sampled data to train models and have achieved remarkable results \cite{END2END_1,DCNet,modl}. However, a key limitation lies in their reliance on large amounts of high-quality fully sampled datasets, the acquisition of which is often time-consuming and challenging for dynamic MRI. Moreover, models trained on specific datasets may exhibit limited generalizability across different scanners or imaging protocols. In contrast, unsupervised learning methods enable training without fully sampled data by either learning the image distribution, such as diffusion model-based reconstructions, or by directly learning from undersampled data. However, most diffusion model-based reconstruction methods still require training data for model learning \cite{liu2024kt, yu2023universal}. Therefore, unsupervised dynamic MRI reconstruction methods in a scan-specific or zero-shot manner, where the network is trained solely on a single undersampled dataset without an external training dataset, are essential for achieving true data-independent reconstruction. Representative unsupervised approaches, such as deep image prior (DIP) \cite{ulyanov2018deep} and ConvDecoder\cite{ConvDecoder}, utilize scan-specific convolutional neural networks (CNNs) as convolutional image priors, mapping a parameter space to an image space while providing implicit regularizations. However, the discrete nature of CNNs limits their ability to comprehensively capture the continuity of imaging objects.

Recently, implicit neural representation (INR) has emerged as a scan-specific DL paradigm for accelerated MRI \cite{shen2022nerp, feng2023imjense, catalan2025unsupervised, zhu2025implicit}. Specifically, INR models signal as a continuous function over spatial coordinates using a neural network, which is trained exclusively on the current scan data by enforcing data consistency between its output and the acquired $k$-space data. To improve reconstruction quality, regularizations derived from CS-based reconstruction methods are often incorporated into INR models as loss terms \cite{ feng2023imjense, CineJENSE, k-t, catalan2025unsupervised, shaoSTINR, feng2025spatiotemporal}. In dynamic MRI, INR has been applied to directly represent signals in the $k$-space domain \cite{huang2023neural,kunz2024implicit}, but such approaches often struggle to achieve stable reconstructions due to the inherent complexity of $k$-space signals. An alternative strategy is to model dynamic images using INR in the spatial and temporal domains, combined with explicit sparse or low-rank regularizations \cite{CineJENSE,feng2025spatiotemporal,k-t,shaoSTINR,catalan2025unsupervised}. Beyond representing the images themselves, INR has also been employed to model their subspace components, further improving reconstruction accuracy \cite{huang2024subspace,zhao2012image,shen2025imj}. Although these methods have achieved impressive reconstruction performance, they still face challenges when applied to highly undersampled data. The main reason is representing the intricate spatiotemporal variations in dynamic MRI using INR requires a large number of network parameters, while the limited data constraints under high undersampling make it difficult to effectively train such high-capacity models, leading to suboptimal reconstructions.

To address these challenges, we move beyond INR parameterization of the entire components and instead use INR to parameterize factor matrices in the tensor decomposition. This design preserves the benefits of continuous representations while compressing the parameter space by expressing the data through learnable tensor function. Specifically, dynamic image series are organized into high-dimensional tensors, which are factorized into a core tensor and several factor matrices. Each matrix is represented by INR as a continuous factor function, allowing the tensor to be modeled as a learnable tensor function. To further enhance representation efficiency and multidimensional low-rankness, we group image patches into non-local tensors to jointly capture temporal correlations and inter-patch similarities, instead of treating the entire image series as a global tensor. Moreover, spatiotemporal TV and low-rank regularizations are incorporated into the loss function to improve reconstruction quality. The framework is trained solely from a single undersampled $k$-space dataset, enabling scan-specific adaptation. Experimental results demonstrate that TenF-INR achieves a high acceleration rate up to 21-fold and consistently outperforms state-of-the-art supervised and unsupervised methods. The major contributions are as follows:
\par (1) We introduce TenF-INR, a novel unsupervised framework which leverages INR to model tensor functions within tensor decomposition, thereby enabling continuous and expressive MR image representation beyond conventional discrete bases.
\par (2) We propose a patch-based non-local tensor modeling strategy that integrates INR-based functional representations with low-rank tensor decomposition. By jointly exploiting temporal correlations and inter-patch similarities, TenF-INR enforces multidimensional low-rankness more effectively than global tensor models, offering a new paradigm for dynamic MRI reconstruction.
\par (3) We demonstrate that TenF-INR achieves efficient reconstruction from highly undersampled $k$-space data, supporting acceleration factors up to 21. It not only surpasses state-of-the-art unsupervised deep learning and traditional methods but also outperforms supervised approaches, highlighting its robustness and generalization ability.


\begin{figure*}[!htbp]
    \centering
    \includegraphics[width=\textwidth]{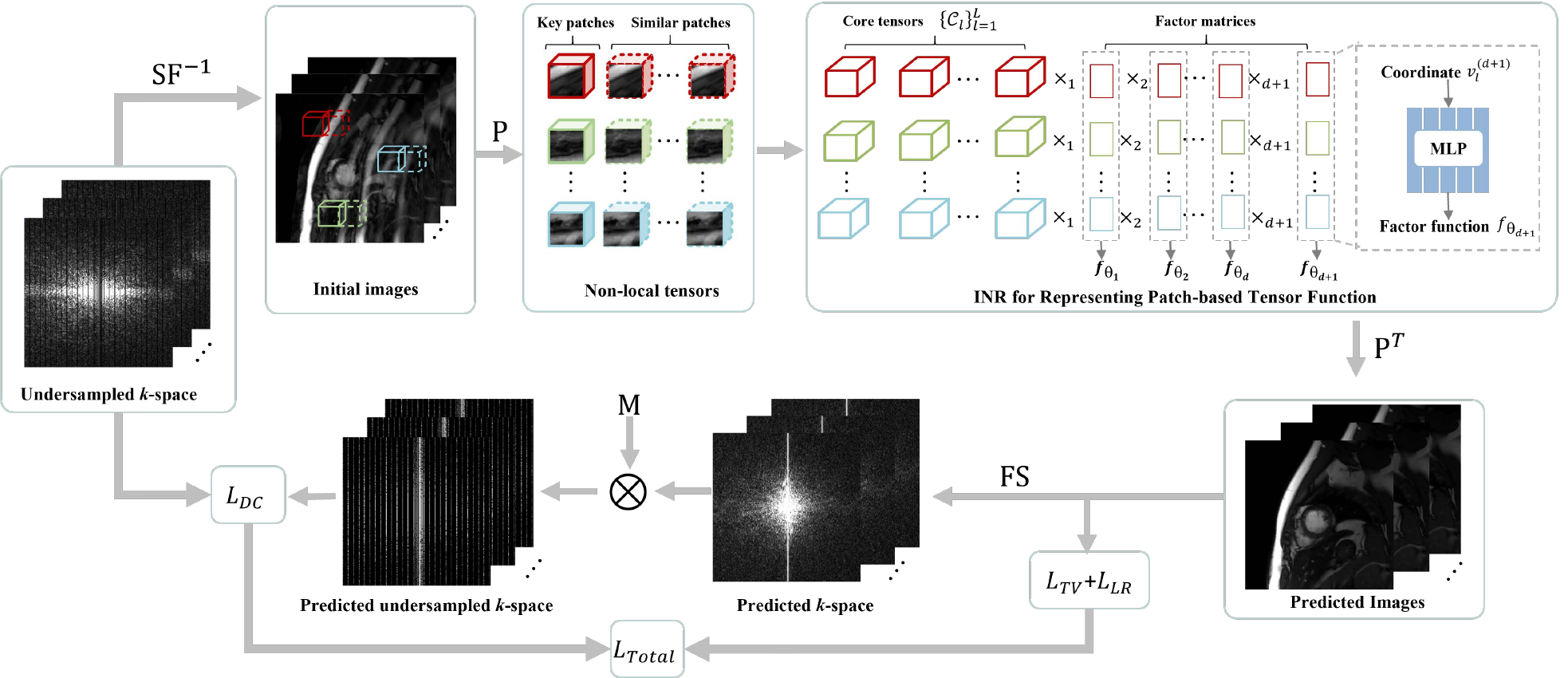}
    \caption{The framework diagram of TenF-INR. ${\mathbf{F}}$ represents the Fourier transform operator, $\mathrm{\mathbf{S}}$ represents an operator which multiplies the sensitivity map coil-by-coil, $\mathrm{\mathbf{F}}^{-1}$ represents the inverse Fourier transform operator, $\mathrm{\mathbf{M}}$ denotes the undersampling mask, and the operators $\mathbf P$ and $\mathbf P^{T}$ represent the similar patch extraction operation and its adjoint operation, respectively.}
    \label{fig1:overall framework}
\end{figure*}

\section{Background}

\subsection{Problem Formulation in Dynamic MRI Reconstruction}
The forward model of dynamic MRI reconstruction can be formulated as:
\begin{equation}
\label{eq:forward_model}
\mathbf {Y} = \mathbf {AX} + \epsilon 
\end{equation}
where $\mathbf X \in \mathbb{C}^{N_x \times N_y \times N_t}$ represents the dynamic MR images to be reconstructed, $\mathbf Y \in \mathbb{C}^{N_x \times N_y \times N_t \times N_s}$ is the observed $k$-space data, $N_x$ and $N_y$ denote the number of frequency and phase encoding lines, respectively, and $N_t$ denotes the frame number, $N_s$ denotes coil number. The $\mathbf{A} = \mathbf{MFS}$ is the encoding operator, $\mathbf M$ is the undersampling matrix, $\mathbf F$ is the Fourier transform operator, and $\mathbf S$ denotes the operator that multiplies the sensitivity map coil-by-coil, $\epsilon$ denotes the measurement noise. The image reconstruction can be formulated as a regularized optimization problem:
\label{optimization_problem equation}
\begin{equation}
  \begin{gathered}
\underset{\mathbf X}{\operatorname{\arg \min}} \|\mathbf Y - \mathbf{A \mathbf{X}}\|_F^2 + \lambda R(\mathbf{X}) 
\end{gathered}  
\end{equation}
where $R(\mathbf{X})$ is the regularization term and $\lambda$ is the weighting parameter.

\subsection{Tensor Decomposition}
Tensor decomposition provides an effective framework to analyze multidimensional data. Classical decomposition methods, such as Tucker and canonical Polyadic (CP) decompositions, factorize a tensor into multiple mode-wise components, capturing correlations across different dimensions \cite{tucktensor}. In this study, Tucker decomposition is employed for dynamic MRI modeling.

\par Let tensor  \(\mathcal{X} \in \mathbb{C}^{N_1 \times N_2 \times \cdots \times N_d}\) denote a $d$-dimensional array used to represent the observed data. The mode-\(i\) unfolding of the tensor \(\mathcal{X}\) is defined as $\mathbf{X}^{(i)} \in \mathbb{C}^{\,N_i \times \big(\prod_{j \ne i}^{d} N_j\big)}$ arranging the data along the \(i\)-th dimension as the columns of \(\mathbf{X}^{(i)}\). The reverse operation, folding, rearranges the elements of \(\mathbf{X}^{(i)}\) into the $d$-dimensional tensor \(\mathcal{X}\). According to Tucker decomposition, a tensor can be decomposed into a core tensor multiplied with multiple factor matrices along each mode with the following formula:
\begin{equation}
\label{eq:tucker_decomposition} 
\mathcal{X} = \mathcal{C} \times_1 \mathbf{U}^{(1)} \times_2 \cdots \times_d \mathbf{U}^{(d)}
\end{equation}
where $\mathcal{C} \in \mathbb{C}^{r_1 \times \cdots \times r_d}(r_i \leq N_i, i = 1, 2, ..., d)$ is the core tensor, $r_i$ is the rank of $\mathbf{X}^{(i)}$, \(\{\mathbf{U}^{(i)} \in \mathbb{C}^{N_i \times r_i}\}_{i=1}^d\) is the factor matrices along the $i$-th mode. Here, the core tensor can be treated as the tensor weights when integrating those factors in different modes.

\subsection{INR for Representating Dynamic MR Images}
INR enables dynamic MR images to be modeled as a continuous function parameterized by a neural network $f_{\theta}$. This network takes  spatiotemporal coordinates as input and is typically implemented using a multi-layer perceptron (MLP). The mapping can be defined as:
\begin{equation}
f_\theta: (x,y,t) \rightarrow \boldsymbol{I} \quad \text { with } (x,y,t) \in \mathbb{R}^3, \boldsymbol{I} \in \mathbb{C}
\end{equation}
where $(x,y,t)$ denotes the spatiotemporal coordinate space, $\boldsymbol{I}$ represents the corresponding complex signal value space, and $\theta$ is the network parameters. Previous studies have demonstrated that employing periodic activation functions, such as the sine activation function \cite{sitzmann2020implicit}, or encoding input coordinates into a high-dimensional space using techniques like positional encoding \cite{INR_2}, Fourier feature mapping \cite{tancik2020fourier}, and hash encoding \cite{hash}, enhances the MLP's capacity to capture high-frequency information. In this study, rather than using INR to directly represent the dynamic MR images, we employ INR to model the tensor function of the low-rank tensor decomposition, with details elaborated in the subsequent sections.

\section{Method}
\subsection{INR for Representating Tensor Function}
\par To further generalize Tucker decomposition from discrete data to continuous domains, the concept of low-rank tensor functions is introduced. Instead of working with tensor entries indexed by discrete coordinates, a tensor function models a continuous mapping \(f(\cdot):\Omega_{{obs}} \subset \mathbb{C}^{d} \to\mathbb{C}\) using a set of continuous factor functions \(f_{\theta_{i}}\), where \(\Omega_{obs}\) denotes the observation domain. For any coordinate vector variables \(\mathbf{v}=(v^{(1)}, v^{(2)}, \ldots, v^{(d)})\subset \Omega_{obs}\), the tensor function can be decomposed as the product of a core tensor \(\mathcal{C}\) and \(d\) mode-wise factor functions in a Tucker-like format:
\begin{equation}
\label{eq:tensor_function_decomposition} 
f(\mathbf{v}) = \mathcal{C} \times f_{\theta_{1}}(v^{(1)}) \times_2 \cdots \times_d f_{\theta_{d}}(v^{(d)})
\end{equation}

Early works used predefined functional bases, such as Gaussian \cite{yokota2015smooth}, polynomial \cite{debals2017nonnegative}, Fourier \cite{kargas2021supervised}, or Chebyshev expansions \cite{hashemi2017chebfun}, as the factor functions. However, such fixed functional bases often lack the flexibility to accurately capture the complex and data-adaptive variations commonly found in real-world signals. In this study, we utilize the continuous nature and expressive power of INRs to model the factor functions, thereby representing the similar patches more effectively as a tensor function in the proposed TenF-INR method. 

\subsection{INR for Representing Patch-based Tensor Function}
In this study, an unsupervised patch-based reconstruction method using learnable tensor function with INR for accelerated dynamic MRI is proposed. Fig.~\ref{fig1:overall framework} illustrates the flow diagram of the proposed TenF-INR method. First, given the dynamic image \(\mathbf{X} \in \mathbb{C}^{N_x \times N_y \times N_t}\) to be reconstructed, nearest similar patches with strong correlations are extracted using a patch selection operator \(\mathbf P\). Each group of nearest similar patches is then organized into a non-local tensor. Second, each non-local tensor is decomposed via Tucker decomposition in (\ref{eq:tensor_function_decomposition}) into a core tensor and a set of factor matrices \(\{\mathbf{U}^{(i)}\}_{i=1}^{d+1}\), where \(d=4\) corresponds to the spatial and temporal dimensions, as well as real-imaginary channel of \(\mathbf{X}\) in this study. The core tensor is parameterized as \(\{\mathbf{C}_l\}_{l=1}^L\), where \(L\) denotes the number of patch groups, and is updated during training. The factor matrices are continuously modeled using independent INRs \(\{f_{\theta_i}(\cdot)\}_{i=1}^{d+1}\), which act as factor functions in a continuous domain. As a result, each non-local tensor becomes a learnable tensor function, and can be optimized through network training. Finally, by querying the learned tensor functions at the appropriate coordinates, the reconstructed image \(\mathbf{X}\) is recovered via the inverse patch selection operator \(\mathbf P^T\). The entire framework is trained based on data consistency, integrated with composite regularizations that will be detailed in the subsequent section.
\subsubsection{Non-local Tensor Construction}
\label{sec:partitioning_patch_construction}
\begin{figure}[!htbp]
\centering
\includegraphics[width=\columnwidth]{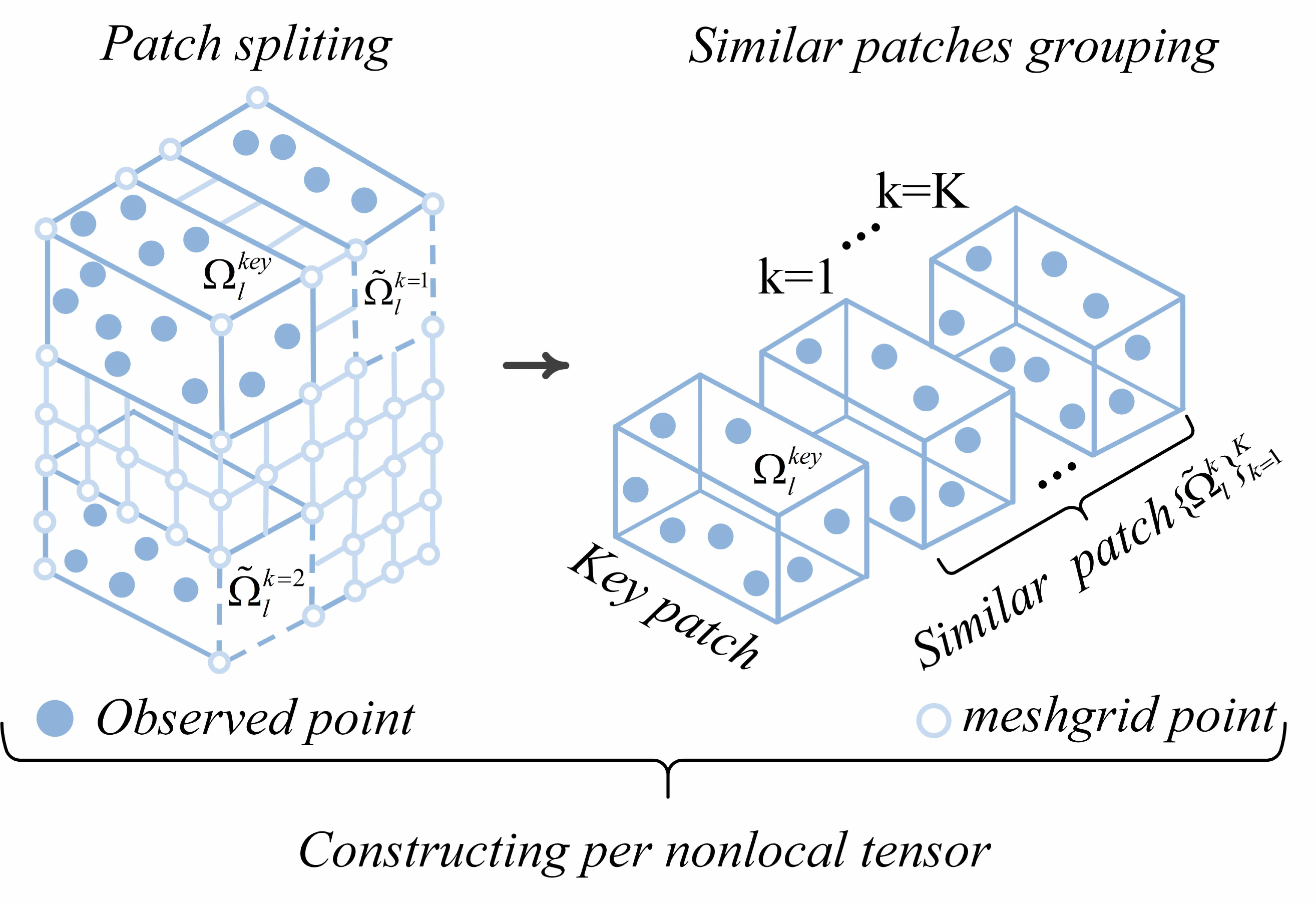}
\caption{Illustrations of the non-local tensor construction process in a three-dimensional scenario with unit numbers $n_1=6, n_2 = n_3 = 5$ and patch size $p = 2$. Here, $\Omega_l^{\text{key}}$ indicates a key patch, and $\tilde\Omega_l^{k=1}$, $\tilde\Omega_l^{k=2}$ are the two nearest similar patches extracted from the $l$-th key patch $\Omega_l^{\text{key}}$.}
\label{grouping process}
\end{figure}

In this section, we first split the image into several non-overlapping key patches. Specifically, given the initial image $\mathbf X_{init} \in \mathbb{C}^{N_x \times N_y \times N_t}$, which is the zero-filling image from undersampled data, and the patch size $p$, we construct $L = (N_x/p)(N_y/p)$  patches with stride $p$ along the spatial dimension (frequency encoding, phase encoding directions)\footnote{In the case where $N_x$, $N_y$ is not divisible by $p$, we use replication padding to expand data boundaries such that  $N_x$, $N_y$ is divisible by $p$.}. For each patch, the corresponding pixels across all $N_t$ temporal frames are included, forming a 3D key patch containing $p^2\times N_t$ basic units. Second,  we extract $K$ similar patches $\{\tilde\Omega_l^{k}\}_{k=1}^{K}$ for each key patch $\{{\Omega_l^{\text{key}}}\}_{l=1}^{L}$ using the block matching method in \cite{dabov2007image}, which computes the Euclidean distance between each key patch and its neighboring patches to select the most similar $K$ neighboring patches based on these distances. Third, the similar patches are grouped together to a 4D tensor, and considering the real and imaginary components of MRI image,  a 5D non-local tensor $\mathcal{X}_l \in \mathbb{R}^{n_1 \times n_2 \times \cdots \times n_{d+1}} (d = 4)$ can be obtained, where the first three dimensions represent the frequency encoding, phase encoding and temporal direction, and the last dimension corresponds to the similar patch dimension. The whole process can be expressed as an operator $\mathbf P$, and Fig.~\ref{grouping process} shows an overview of the process for constructing the non-local tensor.

\subsubsection{Reconstruction Model of TenF-INR}
For each non-local tensor, the core tensor is parameterized as $\mathcal{C}_l \in \mathbb{R}^{r_1 \times \cdots \times r_{d+1}}$, and the corresponding factor matrices are continuously modeled by independent INRs \(\{ f_{\theta_i}(\cdot)\}_{i=1}^{d+1} \) over a ($d+1$)-dimensional  coordinate vector 
$\mathbf{v}_l = (v_l^{(1)}, v_l^{(2)}, \ldots, v_l^{(d+1)})$
, where $\mathbf{v}_l \in \mathbb{R}^{d+1}$. Considering computational efficiency, \(\ f_{\theta_i}(\cdot)\ \) shares parameters across different non-local tensors. The resulting tensor function can be derived from (\ref{eq:tensor_function_decomposition}) as:
\begin{equation}
f_{\theta_l, \mathcal C_l}(\mathbf{v}_l) = \mathcal C_l \times_1 f_{\theta_1}(v_l^{(1)}) \times_2 \cdots \times_{d+1} f_{\theta_{d+1}}(v_l^{(d+1)})
\label{Parameterised tensor function decomposition}
\end{equation}

Incorporating the learnable tensor function into the reconstruction model of TenF-INR, the reconstruction formula can be expressed as:
\begin{equation}
\begin{gathered}
\underset{\{\mathcal{C}_{l}\}_{l=1}^L, \{\theta_{i}\}_{i=1}^{d+1}}{\arg\min} \ \left\| \mathbf{Y} - \mathbf{A}\mathbf{X} \right\|_F^2 + \lambda R(\mathbf{X}) \\
\text{s.t.} \quad \mathbf{X} = \sum_{l=1}^L \left[ P^T f_{\theta}(\mathbf{v}_l) \right], \\
f_{\theta, \mathcal{C}_l}(\mathbf{v}_l) = \mathcal{C}_l \times_1 f_{\theta_1}(v_l^{(1)}) \times_2 \cdots \times_{d+1} f_{\theta_{d+1}}(v_l^{(d+1)}) \label{eq:TenF-INR_optimization}
\end{gathered}
\end{equation}
where $\mathbf P$  denotes the patch selection operator, which extracts similar patches to construct per non-local tensor, and $\mathbf P^T$ denotes the adjoint operation, which places the patches back to their original spatial locations in the image. Two constraints are implemented to improve the reconstruction performance. One is the spatiotemporal TV constraint, and the other is the low-rank constraint of the Casorati matrix of image \(\mathbf{X}\).  

\subsubsection{Loss Function}
The network can be trained using the following loss function:
\begin{align}
L_{\text{total}} &= \underbrace{\ \left\| \mathbf{Y} - \mathbf{A}\mathbf{X} \right\|_F^2}_{L_{\text{DC}}} 
 + \underbrace{\lambda_S TV(\mathbf X)}_{L_{\text{TV}}} 
+ \underbrace{\lambda_L \|\mathbf{C} (\mathbf X)\|_*}_{L_{\text{LR}}}
\label{eq:loss_total}
\end{align}
where $\mathbf{C} (\mathbf{X})$ demotes the Casorati matrix of $\mathbf X$, which is formed by vectorizing each frame of $\mathbf X$ into a column vector of a matrix. In the loss function, the first term \(L_\text{DC}\)  ensures data consistency with the acquired $k$-space data. The second term \(L_\text{TV}\)  represents the loss for TV constraint and the last low-rank term \(L_\text{LR}\)  enhances the low-rankness in temporal direction for dynamic MRI.  \( \lambda_S \) and \( \lambda_L \) are regularization parameters that balance the contributions of \(L_\text{TV}\) and \(L_\text{LR}\) terms, respectively. Once the network is trained, an additional step is introduced to refine the reconstruction by replacing the predicted $k$-space with the acquired $k$-space data at data sampling locations, yielding the final reconstructed image~\cite{feng2023imjense}.
\section{Experiments}
\label{sec:experiments}
\subsection{Datasets }
Three datasets were used in this study: an in-house cardiac cine dataset, an in-house dynamic abdomen imaging dataset, and a publicly available cardiac cine dataset (OCMR) \cite{chen2020ocmr}.
The in-house cine dataset was mainly used for retrospective experiments to demonstrate the effectiveness of the proposed TenF-INR method, whereas the abdomen and OCMR datasets were utilized for prospective experiments to evaluate generalization.

\label{subsec:datasets}

\subsubsection{In-house cine dataset} 
\begin{figure}[htbp]
    \centering
    \includegraphics[width=0.8\linewidth]{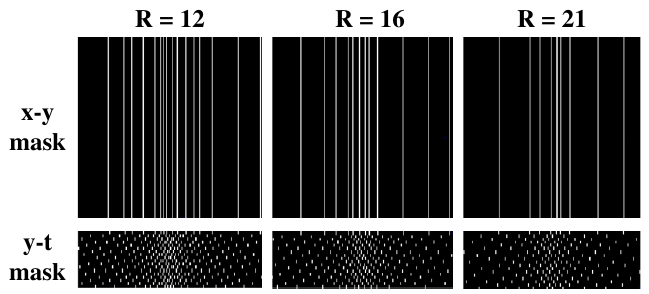}
    \caption{Examples of the undersampling patterns for acceleration factors R = 12, 16, and 21. Sampled $k$-space points are marked in white, with the unacquired $k$-space regions shown in black.}
    \label{fig:sampling_patterns}
\end{figure}

The fully sampled data were collected from 29 healthy volunteers on a 3T scanner (MAGNETOM Trio, Siemens Healthcare, Erlangen, Germany) with a 20-channel receiver coil. For each volunteer, 10 to 13 short-axis slices were imaged using a segmented balanced steady-state free precession (bSSFP) sequence during breath-holding, resulting in a total of 386 slices collected. Imaging parameters were: acquisition matrix = $256 \times 256$,  slice thickness = 6 mm, TE/TR = 1.5/3.0 ms. The temporal resolution was 40 ms and reconstructed to produce 25 phases to cover the entire cardiac cycle. For supervised learning, data augmentation was applied via stride and cropping, using a $192 \times 192 \times 18$ $(x \times y \times t)$ sliding box with strides of 25, 25, and 7 along the $x$, $y$, and $t$ directions, respectively. The fully sampled $k$-space data were retrospectively undersampled by variable density incoherent spatiotemporal acquisition sampling masks \cite{rich2020cartesian} with acceleration factors (R) of 12, 16, and 21. The undersampling patterns are shown in Fig.\ref{fig:sampling_patterns}.

\subsubsection{In-house dynamic abdomen dataset} The prospective dynamic abdomen dataset was acquired using a 3T MRI scanner (uMR790, United Imaging Healthcare, Shanghai, China) using a 12-channel receiver coil and a 2D fast spin echo (FSE) sequence. Imaging parameters were: acquisition matrix = $340 \times 340$, slice thickness = 3 mm, TE/TR = 119.88/2000 ms. The 7-fold undersampled data with a pseudo-random trajectory were collected in real-time mode under breath-holding, capturing 6 phases per slice.
\subsubsection{The OCMR dataset} The prospective OCMR dataset \cite{chen2020ocmr} was acquired using a 3T MRI scanner (MAGNETOM Prisma, Siemens Healthcare, Erlangen, Germany) with a bSSFP sequence and a 34-channel receiver coil array. Imaging parameters were: acquisition matrix = 192 × 144, slice thickness = 8 mm, and TE/TR = 1.05/2.4 ms. The 9-fold undersampled data with a pseudo-random trajectory were collected in real-time mode under free-breathing, capturing 65 frames per slice. Coil sensitivity maps for both retrospective and prospective data were estimated using the ESPIRiT method \cite{ESPIRiT}, based on the time-averaged $k$-space center extracted from the undersampled data.

\subsection{Implementation Details}
\label{subsec:Implementation Details}
In this study, an MLP with one input layer, one hidden layer, and one output layer was employed to parameterize each factor function. The hidden layer comprises 126 neurons, and all layers, except the final one, utilize the periodic activation function \cite{INR_1}. The weights of the first layer was initialized with $U[-1/n_l, 1/n_l]$ and other layers were initialized with a uniform distribution $U[-\sqrt{6}/n_l, \sqrt{6}/n_l]$, where $n_l$ represents the number of input features for each layer. 

To optimize performance, hyperparameter tuning was conducted using the Ray Tune tool~\cite{Ray_tune}. The rank of the Tucker decomposition was set to $(r_1, r_2, r_3, r_4, r_5) = (2, 2, 16, 2, 5)$, with the patch size $p=2$ and the number of similar patches $K=20$. The ADAM optimizer was used with an initial learning rate of 0.0001, which was reduced by 80\% for every 500 iterations. A weight decay parameter of 0.38 was applied, and the regularization parameters $\lambda_{S}$  and $\lambda_{L}$ were set to $1 \times 10^{-3}$ and $5 \times 10^{-6}$, respectively. The total number of iterations was set to 12000. 

All experiments were conducted on an Ubuntu 20.04.4 operating system equipped with an NVIDIA A100 Tensor Core GPU (80 GB memory), using the PyTorch 2.5.0 framework with CUDA 11.8 and cuDNN support.

\begin{table}[!htbp]
\caption{Comparisons of different methods for the retrospective reconstructions with different acceleration factors (AF). The best results are shown in bold.}
\label{table 1:quantitative metrics}   
\begin{center}
\resizebox{\columnwidth}{!}{%
\setlength{\tabcolsep}{4pt}
\begin{tabular}{cc|ccc}
\hline \hline
  AF                   & Method     & PSNR                & SSIM                   & RMSE                   \\
  \hline \multirow{7}{*}{$\times$12}  
                     & ConvDec & 41.13±0.86          & 0.9619±0.0057          & 0.0079±0.0006          \\
                     & CineJSENSE & 43.96±0.91       & 0.9745±0.0064          & 0.0063±0.0006          \\
                     & MoDL    & 42.64±0.73          & 0.9616±0.0072          & 0.0074±0.0006          \\
                     & Learned DC      & 44.58±1.16          & 0.9809±0.0043          & 0.0059±0.0010          \\
                     & L+S-Net      & 46.87±0.95          & 0.9876±0.0023          & 0.0046±0.0005          \\
                     & TenF-INR       & \textbf{51.21±0.85} & \textbf{0.9952±0.0007} & \textbf{0.0028±0.0002} \\ 
                     \hline \multirow{7}{*}{$\times$16}  
                     & ConvDec & 40.88±0.60          & 0.9577±0.0058          & 0.0089±0.0006          \\
                     & CineJSENSE & 43.13±0.65       & 0.9702±0.0038          & 0.0069±0.0005          \\
                     & MoDL    & 41.65±0.76          & 0.9562±0.0073          & 0.0088±0.0014          \\
                     & Learned DC      & 43.88±0.99          & 0.9793±0.0028          & 0.0064±0.0007          \\
                     & L+S-Net      & 46.47±0.93          & 0.9867±0.0025          & 0.0048±0.0005          \\
                     & TenF-INR       & \textbf{50.09±0.71} & \textbf{0.9940±0.0009} & \textbf{0.0032±0.0002} \\ 
                     \hline \multirow{7}{*}{$\times$21}
                     & ConvDec & 38.57±1.14          & 0.9393±0.0098          & 0.0119±0.0016          \\
                     & CineJSENSE & 41.74±0.58       & 0.9619±0.0060          & 0.0082±0.0005          \\
                     & MoDL    & 41.06±0.67          & 0.9483±0.0068          & 0.0088±0.0006          \\
                     & Learned DC      & 42.03±1.10          & 0.9696±0.0042          & 0.0080±0.0011          \\
                     & L+S-Net      & 45.49±0.69          & 0.9847±0.0024          & 0.0053±0.0005          \\
                     & TenF-INR      & \textbf{48.62±0.77} & \textbf{0.9922±0.0017} & \textbf{0.0037±0.0003} \\ \hline \hline
\end{tabular}%
}
\end{center}
\end{table}

\subsection{Comparison methods}
TenF-INR was compared with five state-of-the-art methods: two unsupervised approaches-ConvDec \cite{ConvDecoder} and CineJENSE \cite{CineJENSE}- and three supervised approaches-MoDL \cite{modl}, Learned DC \cite{DCNet}, and L+S-Net \cite{L+S-Net}. Among the unsupervised methods, ConvDec employs a DIP-based strategy, while CineJENSE adopts INR for joint image and sensitivity map reconstruction. For the supervised approaches, MoDL introduces a model-based reconstruction framework with CNN-based regularization priors, Learned DC integrates data consistency into a model-driven unrolled deep learning framework for dynamic reconstruction, and L+S-Net enhances reconstruction quality by incorporating learned singular value thresholding within a low-rank plus sparse decomposition model.
Reconstruction performance was assessed through visual comparisons and quantitative metrics, including Peak Signal-to-Noise Ratio (PSNR), Structural Similarity Index (SSIM), and Root Mean Square Error (RMSE).
Furthermore, considering the data distribution discrepancy, prospective evaluations were performed separately for the cine and abdomen datasets: supervised methods evaluated on the prospective OCMR data were trained on the corresponding fully sampled cine dataset, whereas supervised methods evaluated on the prospective abdomen data were trained on a fully sampled dynamic abdomen dataset \cite{cheng2024dynamic}.

\begin{figure*}[!htbp]
    \centering
    \includegraphics[width=1.85\columnwidth]{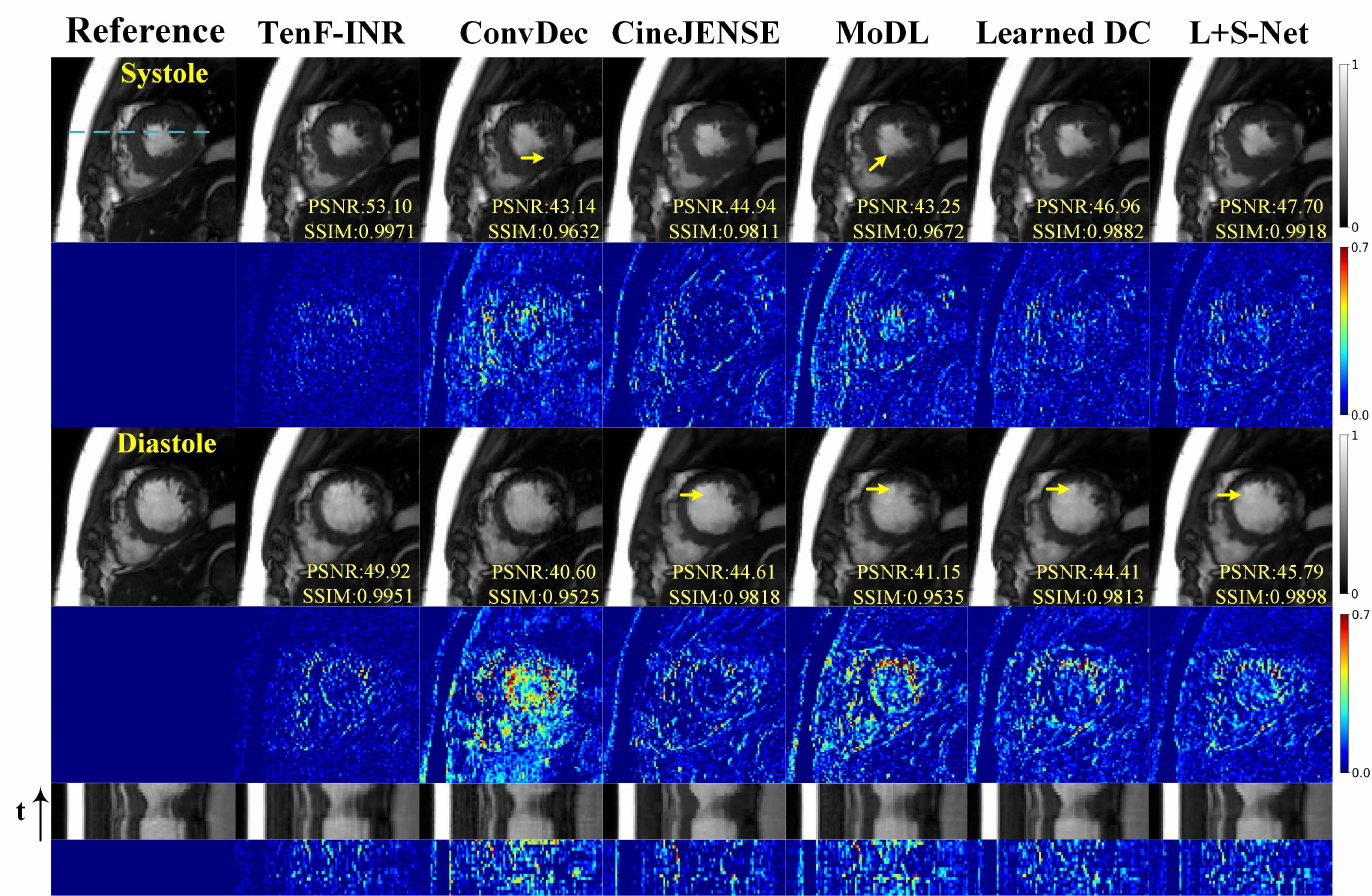}
    \caption{Visual comparison of six reconstruction methods at R = 12. The top four rows show the reconstruction results at systolic and diastolic phases, including PSNR/SSIM metrics and corresponding error maps. The bottom two rows display the y-t dynamic image and its error map, extracted along the blue dashed line. Yellow arrows indicate artifacts or blurs.}
    \label{Results of R = 12}
\end{figure*}

\begin{figure*}[!htbp]
    \centering
    \includegraphics[width=1.85\columnwidth]{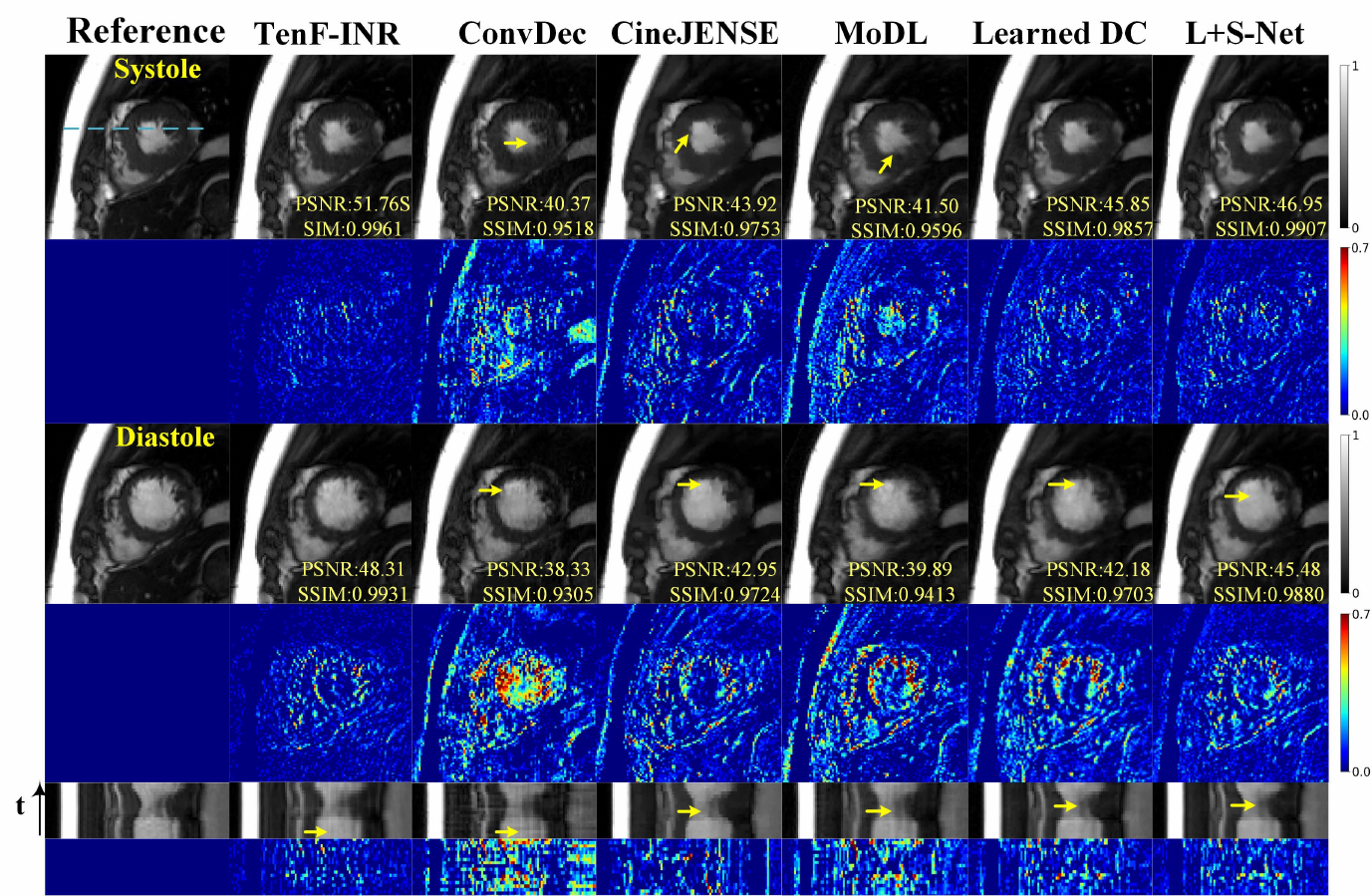}
    \caption{Visual comparison of six reconstruction methods at R = 16. The top four rows show the reconstruction results at systolic and diastolic phases, including PSNR/SSIM metrics and corresponding error maps. The bottom two rows display the y-t dynamic image and its error map, extracted along the blue dashed line. Yellow arrows indicate artifacts or blurs.}
    \label{Results of R = 16}
\end{figure*}

\begin{figure*}[!htbp]
    \centering
    \includegraphics[width=1.85\columnwidth]{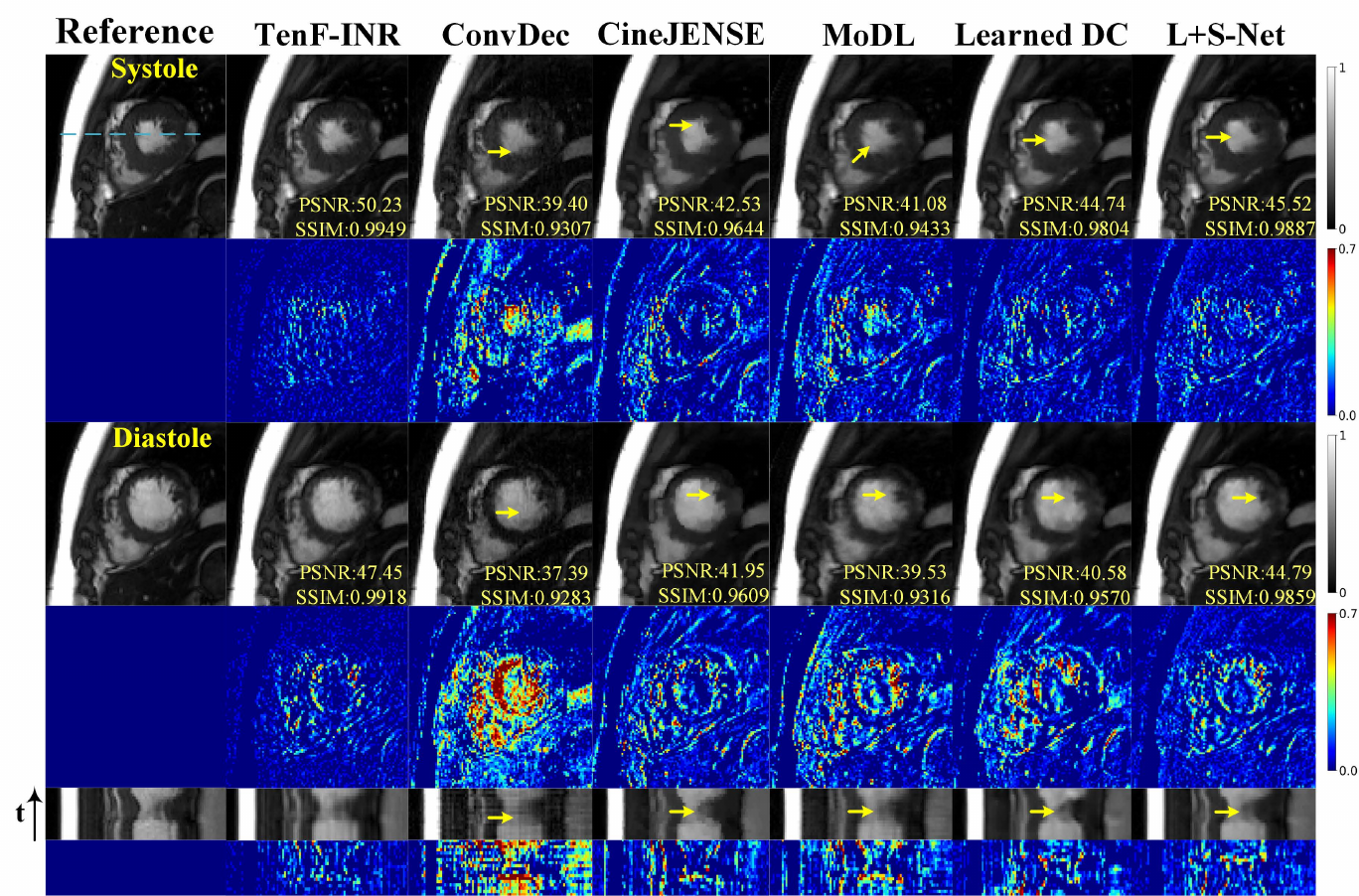}
    \caption{Visual comparison of six reconstruction methods at R = 21. The top four rows show the reconstruction results at systolic and diastolic phases, including PSNR/SSIM metrics and corresponding error maps. The bottom two rows display the y-t dynamic image and its error map, extracted along the blue dashed line. Yellow arrows indicate artifacts or blurs.}
    \label{Results of R = 21}
\end{figure*}

\subsection{Ablation Studies}
In this study, we performed comprehensive ablation experiments to assess the influence of key components within our framework on the reconstruction performance.
\subsubsection{Ablation Study 1} To demonstrate the superiority of the patch-based tensor function representation strategy, an ablation study was conducted to compare the proposed method and the reconstruction framework with global tensor functions (referred to as w/o patch-based) represented by INR using the following formula:
\begin{equation}
\begin{aligned}
&\underset{\mathcal{C},\ \theta}{\arg\min} \left\| \mathbf{Y} - \mathbf{A}f_\theta(\mathbf{v}) \right\|_F^2 +R\left(f_{\theta}(\mathbf{v})\right) \hfill \\
&f_{\theta}(\mathbf{v}) = \mathcal{C} \times_1 f_{\theta_{1}}(v^{(1)}) \times_2 \dots \times_{d+1} f_{\theta_{d}}(v^{(d)}) \hfill
\end{aligned}
\label{eq:global_optimization}
\end{equation}

To ensure a fair comparison, the same regularization terms as the proposed method were used. The Ray Tune tool was also employed to faithfully optimize the parameters for the w/o patch-based method, and the rank of the tensor decomposition was set to \((r_1, r_2, r_3, r_4) = (160, 160, 15, 2)\).
\subsubsection{Ablation Study 2} To evaluate the effectiveness of the two regularization terms, $L_{TV}$ and $L_{LR}$, in enhancing reconstruction performance, we trained the network with specific regularization terms ablated, employing the following loss functions respectively:
\begin{align}
L_{\text{total-1}} &= \left\| \mathbf{Y} - \mathbf{A}\mathbf{X} \right\|_F^2
 + \lambda_S TV(\mathbf X) &&
\label{eq:loss_total_1}
\end{align}

\begin{align}
L_{\text{total-2}} &= \left\| \mathbf{Y} - \mathbf{A}\mathbf{X} \right\|_F^2
 +\lambda_L \|\mathbf{C} (\mathbf X)\|_* &&
\label{eq:loss_total_2}
\end{align}

\begin{align}
L_{\text{total-3}} &= \left\| \mathbf{Y} - \mathbf{A}\mathbf{X} \right\|_F^2 &&
 \label{eq:loss_total_3}
\end{align}

\begin{figure*}[!htbp]
  \centering
  \includegraphics[width=1.85\columnwidth]{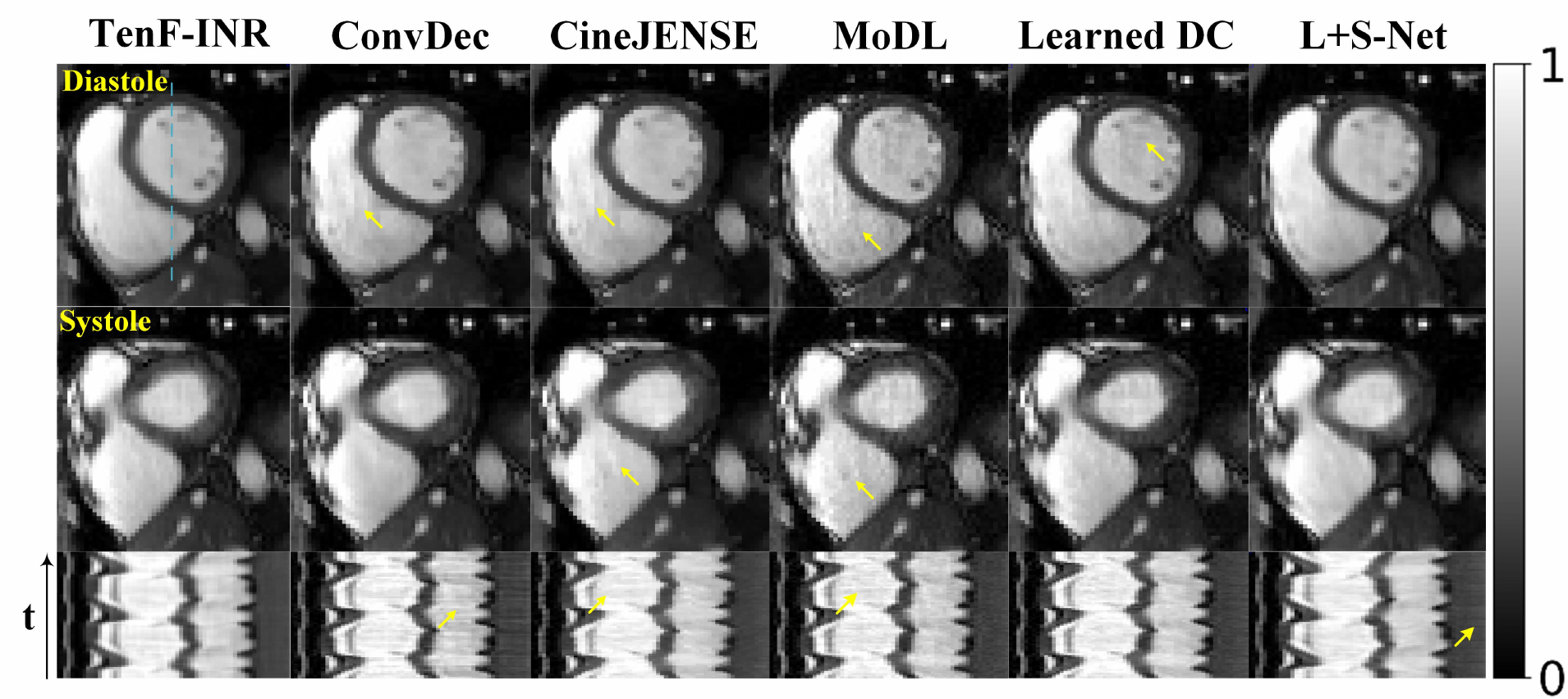}
  \caption{Reconstruction results for prospective dynamic MRI at an acceleration factor of R = 9. The first and second rows show the reconstructed frames at the diastolic and systolic phases, respectively, while the third row illustrates the y–t view along the blue dashed line. Compared with other methods that exhibit noticeable artifacts in the regions indicated by the yellow arrows, TenF-INR produces cleaner and more accurate reconstructions.}
  \label{fig:OCMR_results}
\end{figure*}

\vspace{-1em} 

\begin  {figure*}[!t]
  \centering
  \includegraphics[width=0.85\textwidth]{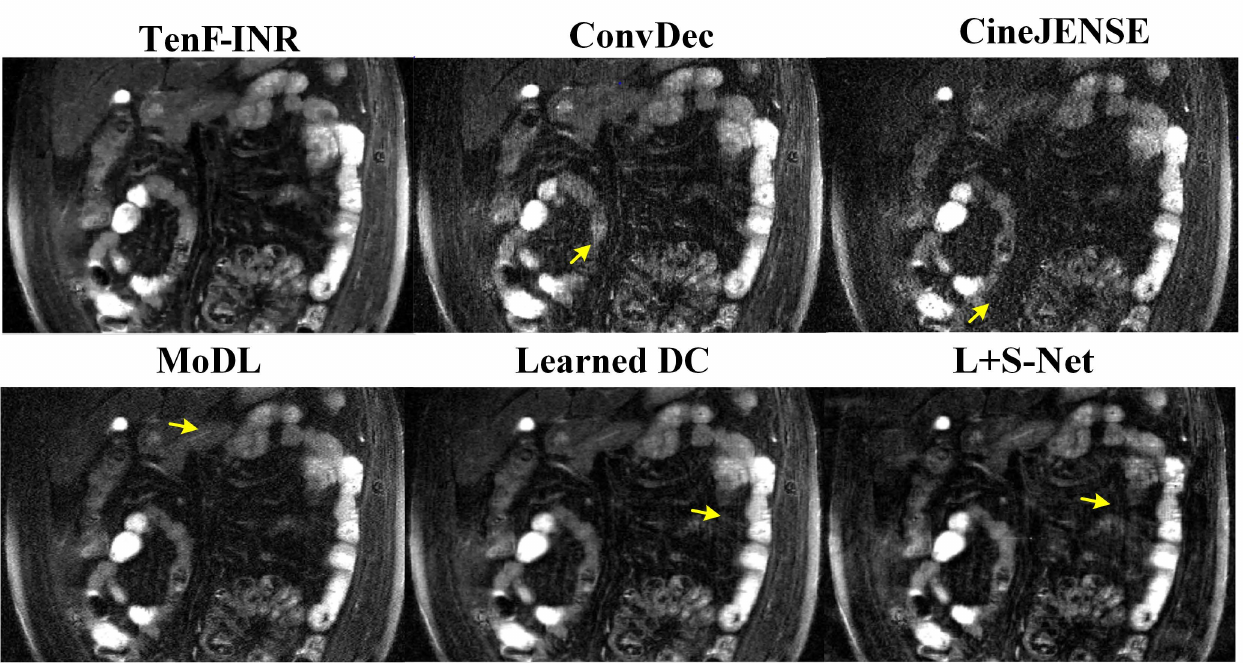}
  \caption{Reconstruction results for the prospective abdomen dataset at R = 7. 
  Compared with other methods that display noticeable artifacts or noise in the regions indicated by the yellow arrows, TenF-INR produces cleaner and more accurate reconstructions.}
  \label{fig:abd_results}
\end{figure*}

\section{Results}
\subsection{Retrospective Reconstruction }
Fig.~\ref{Results of R = 12} shows reconstruction results at diastolic and systolic phases with R = 12. The ConvDec and MoDL reconstructions show noticeable aliasing artifacts in the systolic phase, while CineJENSE, Learned DC, and L+S-Net exhibit blurring in the diastolic phase. Notably, TenF-INR yields high-quality reconstructions with better detail preservation and artifact suppression. Additionally, error images demonstrate that TenF-INR achieves the lowest reconstruction error to the fully sampled reference image.
As depicted in Fig.~\ref{Results of R = 16}, with the increase of acceleration factor, the decline of PSNR and SSIM is observed across all the methods. ConvDec yields the highest error with evident aliasing, while CineJENSE, MoDL, Learned DC, and L+S-Net all suffer from blurring in the papillary muscle region.

TenF-INR maintains a promising image resolution and preserves fine details as the fully sampled image does. Even at a high acceleration factor of R = 21, TenF-INR continues to provide reliable reconstructions, as illustrated in Fig.~\ref{Results of R = 21}.
Table~\ref{table 1:quantitative metrics} presents the quantitative metrics for reconstructions across all compared methods, derived from 25 subjects with a total of 450 images.
TenF-INR consistently outperforms the other methods, achieving the highest PSNR, SSIM, and lowest RMSE values across all acceleration factors. Remarkably, at R = 21, TenF-INR still maintains high quality, with average PSNR of 48.62 dB, which is consistent with the visual results mentioned previously.

\subsection{Prospective Reconstruction}
\subsubsection{The OCMR Dataset}The free-breathing 9-fold prospectively undersampled cardiac dataset was reconstructed using ConvDec, CineJENSE, MoDL, Learned DC, L+S-Net, and TenF-INR.
Fig.~\ref{fig:OCMR_results} illustrates the reconstruction results. Due to the reduced spatial resolution compared to the retrospective study, the image quality of all compared methods is not as good as that observed in the retrospective study.
Nevertheless, TenF-INR outperforms the other five methods, exhibiting fewer artifacts and reduced spatial blurring. In the x--y view, the results from ConvDec, CineJENSE, MoDL, and Learned DC exhibit streaking artifacts (indicated by yellow arrows), while in the y--t view, L+S-Net suffers from noticeable artifacts and the reconstruction of TenF-INR displays sharper edges with reduced noise compared to the other methods, indicating improved accuracy in capturing dynamic information.

\subsubsection{The Abdomen Dataset}
To further evaluate the generalization capability of the TenF-INR method, we compared its reconstructions with those obtained from five state-of-the-art methods on the prospectively undersampled abdomen dataset with R = 7. The reconstruction results are shown in Fig.~\ref{fig:abd_results}. The ConvDec and CineJENSE yeild reconstructions with noticeable noise and artifacts. Although MoDL, Learned DC, and L+S-Net achieve better performance, MoDL still shows noise, while Learned DC and L+S-Net exhibit artifacts, as indicated by the yellow arrows. In contrast, TenF-INR outperforms all compared methods, demonstrating superior noise and artifact suppression.

In the prospective studies, quantitative comparisons are not provided since the ground truth dataset is not available.

\begin{figure*}[!htbp]
\centerline{\includegraphics[width=1.85\columnwidth]{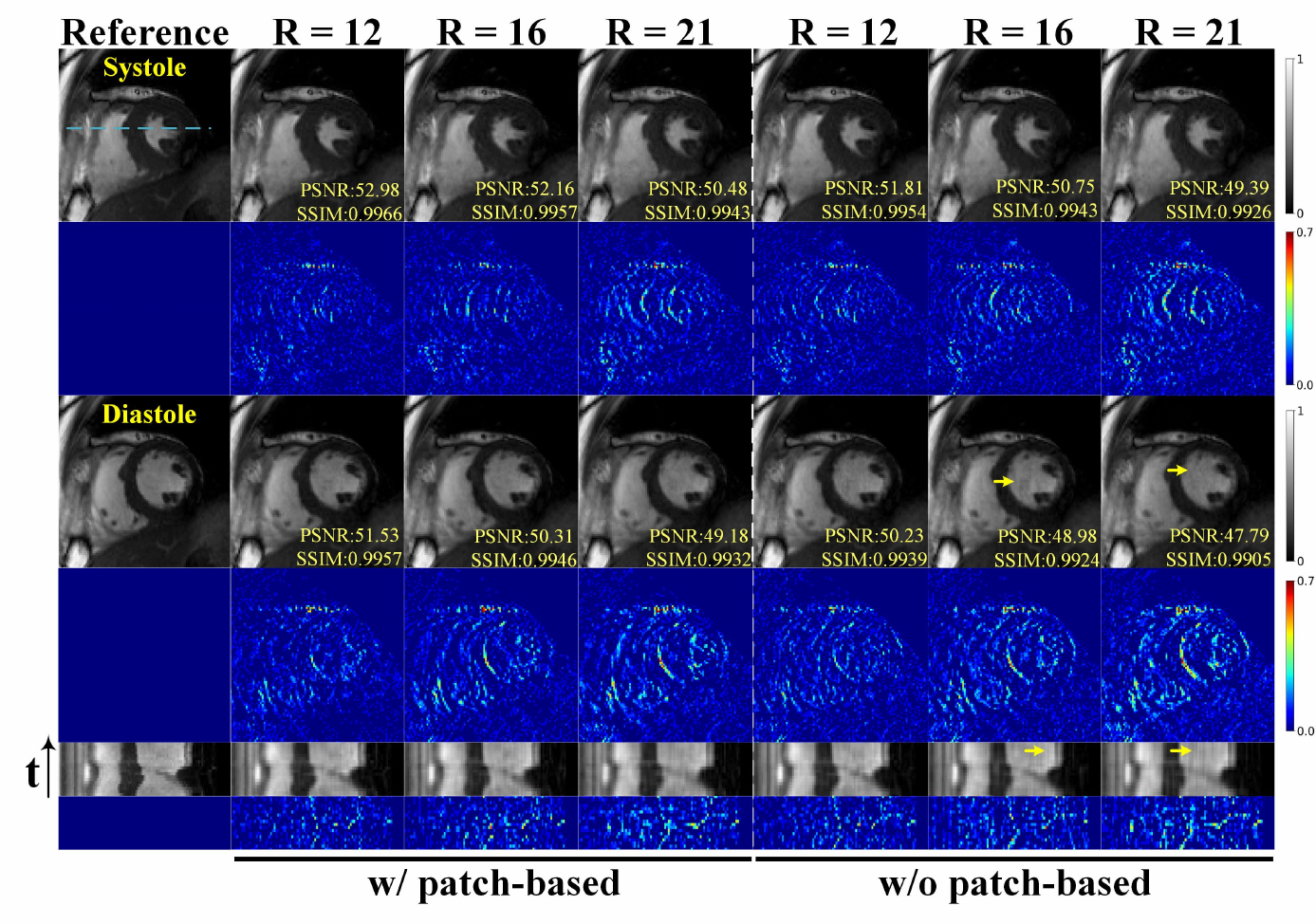}}
\caption{Visual comparison of TenF-INR (w/ patch-based) and global tensor-based INR (w/o patch-based) at  R = 12, 16, 21. The first four rows show reconstructions at diastolic and systolic phases with zoomed-in ROIs, PSNR/SSIM values, and error maps. The bottom two rows display the y-t dynamic images and the error maps, extracted along the blue
dashed line. The global tensor-based method exhibits artifacts and detail loss at  R = 16  and 21 in regions marked by yellow arrows. }
\label{fig:ablation results}
\end{figure*}

\subsection{Results of Ablation Study}

Fig.~\ref{fig:ablation results} shows the reconstruction results for a subject at acceleration factors of R = 12, 16, and 21 in a retrospective study, comparing the TenF-INR method with patch-based processing to the global tensor-based reconstruction method without patch-based processing. Across all acceleration factors, TenF-INR achieves consistently lower reconstruction errors and superior image quality. At R = 12, TenF-INR reconstructs sharper anatomical details with fewer aliasing artifacts than the global tensor-based approach. At R = 16  and R = 21, the global tensor-based method exhibits more streaking artifacts as indicated by the yellow arrows in Fig.~\ref{fig:ablation results}. In contrast, TenF-INR consistently preserves edge clarity and suppresses artifacts across all acceleration factors. The improvement can be attributed to the patch-based approach’s ability to capture both local structures and non-local correlations, enhancing robustness under severe undersampling.
Table~\ref{table of ablation 1&2 TenF-INR}  shows the quantitative results of 450 reconstructed images obtained using patch-based and non-patch-based strategies in Ablation Study 1, further confirming the  advantage of incorporating patch-based processing. Meanwhile, Table~\ref{table of ablation 1&2 TenF-INR} also shows the results of quantitative metrics evaluated under  different regularization terms configurations defined in (\ref{eq:loss_total_1}), (\ref{eq:loss_total_2}), and (\ref{eq:loss_total_3}), respectively. The proposed method consistently yields the highest PSNR and SSIM, and the lowest RMSE, demonstating the superior reconstruction performance by jointly incorporating the $L_{TV}$ and the $L_{LR}$ terms.

\section{Discussion}

In this study, we propose an unsupervised reconstruction method based on learnable low-rank tensor functions using INR for modeling dynamic MR images.
The effectiveness of this function-based factorization in capturing low-rank structures can also be extended to other applications, such as natural image recovery \cite{LRTF,chen2022tensorf}. The proposed TenF-INR method further exploits strong spatiotemporal correlations through patch-based multi-dimensional low-rankness enforced via tensor decomposition, achieving high-quality reconstructions at acceleration factors up to 21 and outperforming state-of-the-art reconstruction methods. Using its unsupervised paradigm, the framework maintains stable performance in different sampling strategies without requiring large fully sampled datasets, which are often difficult to obtain in clinical practice. The compact tensor-function representation also reduces the parameter space and computational burden, facilitating efficient training and deployment, and indicating potential for future clinical practice in dynamic MRI.

\subsection{Optimization of Hyperparameters}
\par In the proposed method, several hyperparameters were optimized to enhance performance, including the tensor decomposition ranks $(r_1, r_2, r_3, r_4, r_5)$, patch size $p$, number of similar patches $K$, regularization parameters $\lambda_{S}$ and $\lambda_{L}$, and sine activation parameter $w$. Following \cite{INR_1}, the parameter $w$ was set to 30, and the other parameters were determined through an optuna search on a randomly selected subset of the whole dataset across different acceleration factors, using the average PSNR between reconstructed and reference images as the evaluation metric. The optimal parameters were then applied to the entire dataset. The search ranges were as follows: $r_1, r_2$, and $p$ varied from 1 to 6 in steps of 1; $r_3$ from 1 to 18 in steps of 1; $r_5$ from 5 to 15 in steps of 5; and $K$ from 10 to 40 in steps of 5, with $r_4$ at 2 due to the fourth dimension representing the real-imaginary channel. 
As for the parameters of regularizers, both $\lambda_{S}$ and $\lambda_{L}$ were optuna-searched within exponential ranges of 1 to ${10}^{-4}$ and ${10}^{-4}$ to ${10}^{-7}$, respectively.

\begin{table*}[!htbp]
\caption{Quantitative evaluation results for Ablation Study 1 and 2 under various acceleration factors, averaged across 450 images from all datasets. Best results are in bold.}
\centering 
\renewcommand{\arraystretch}{1.1}
\setlength{\tabcolsep}{8pt}
\small
\begin{tabular}{ccccccc}
\toprule
\multirow{2}{*}{R} & \multirow{2}{*}{$\mathcal{L}_{\textit{TV}}$} & \multirow{2}{*}{$\mathcal{L}_{\textit{LR}}$} & \multirow{2}{*}{Patch-based} & \multicolumn{3}{c}{Metrics} \\ 
\cmidrule{5-7}
 & & & & PSNR & SSIM & RMSE \\ 
\midrule
\multirow{5}{*}{12 $\times$} 
 & $\times$ & $\checkmark$ & $\checkmark$ &  50.88$\pm$1.88 & 0.9944$\pm$0.0033 & 0.0029$\pm$0.0008  \\
 & $\checkmark$ & $\times$ & $\checkmark$ &  50.48$\pm$1.86 & 0.9942$\pm$0.0036 & 0.0029$\pm$0.0007  \\
 & $\times$ & $\times$ & $\checkmark$ & 48.58$\pm$1.16 & 0.9912$\pm$0.0035 & 0.0038$\pm$0.0006  \\
 & $\checkmark$ & $\checkmark$ & $\times$ &49.56$\pm$1.87& 0.9929$\pm$0.0034  & 0.0027$\pm$0.0007  \\
 & $\checkmark$ & $\checkmark$ & $\checkmark$ &\textbf{51.21$\pm$0.85} & \textbf{0.9952$\pm$0.0010} & \textbf{0.0028$\pm$0.0002} \\
\midrule
\multirow{5}{*}{16 $\times$} 
 & $\times$ & $\checkmark$ & $\checkmark$ & 49.60$\pm$1.78 & 0.9934$\pm$0.0037 & 0.0034$\pm$0.0009  \\
 & $\checkmark$ & $\times$ &  $\checkmark$ & 49.29$\pm$1.76 & 0.9930$\pm$0.0040 & 0.0033$\pm$0.0008  \\
 & $\times$ & $\times$ & $\checkmark$ & 47.34$\pm$1.15 & 0.9890$\pm$0.0040 & 0.0044$\pm$0.0007 \\
 & $\checkmark$ & $\checkmark$ & $\times$ &48.14$\pm$1.75  &0.9915$\pm$0.0038  &0.0040$\pm$0.0009  \\
 & $\checkmark$ & $\checkmark$ &  $\checkmark$ & \textbf{50.09$\pm$0.71} & \textbf{0.9947$\pm$0.0010} & \textbf{0.0032$\pm$0.0002} \\
\midrule
\multirow{5}{*}{21 $\times$} 
 & $\times$ & $\checkmark$ & $\checkmark$ & 48.01$\pm$1.66 & 0.9912$\pm$0.0045 & 0.0039$\pm$0.0009  \\
 & $\checkmark$ & $\times$ & $\checkmark$  & 47.10$\pm$1.57 & 0.9901$\pm$0.0043 & 0.0042$\pm$0.0010  \\
 & $\times$ & $\times$ &  $\checkmark$ & 44.84$\pm$1.01 & 0.9834$\pm$0.0046 & 0.0058$\pm$0.0007  \\
 & $\checkmark$ & $\checkmark$ &  $\times$ &46.89$\pm$1.60  &0.9888$\pm$0.0066  &0.0040$\pm$0.0010  \\
 & $\checkmark$ & $\checkmark$ & $\checkmark$ & \textbf{48.62$\pm$0.77} & \textbf{0.9922$\pm$0.0017} & \textbf{0.0037$\pm$0.0003} \\
\bottomrule
\end{tabular}
\label{table of ablation 1&2 TenF-INR}
\end{table*}

\begin{figure*}[!htbp]
    \centerline{\includegraphics[width=1.85\columnwidth]{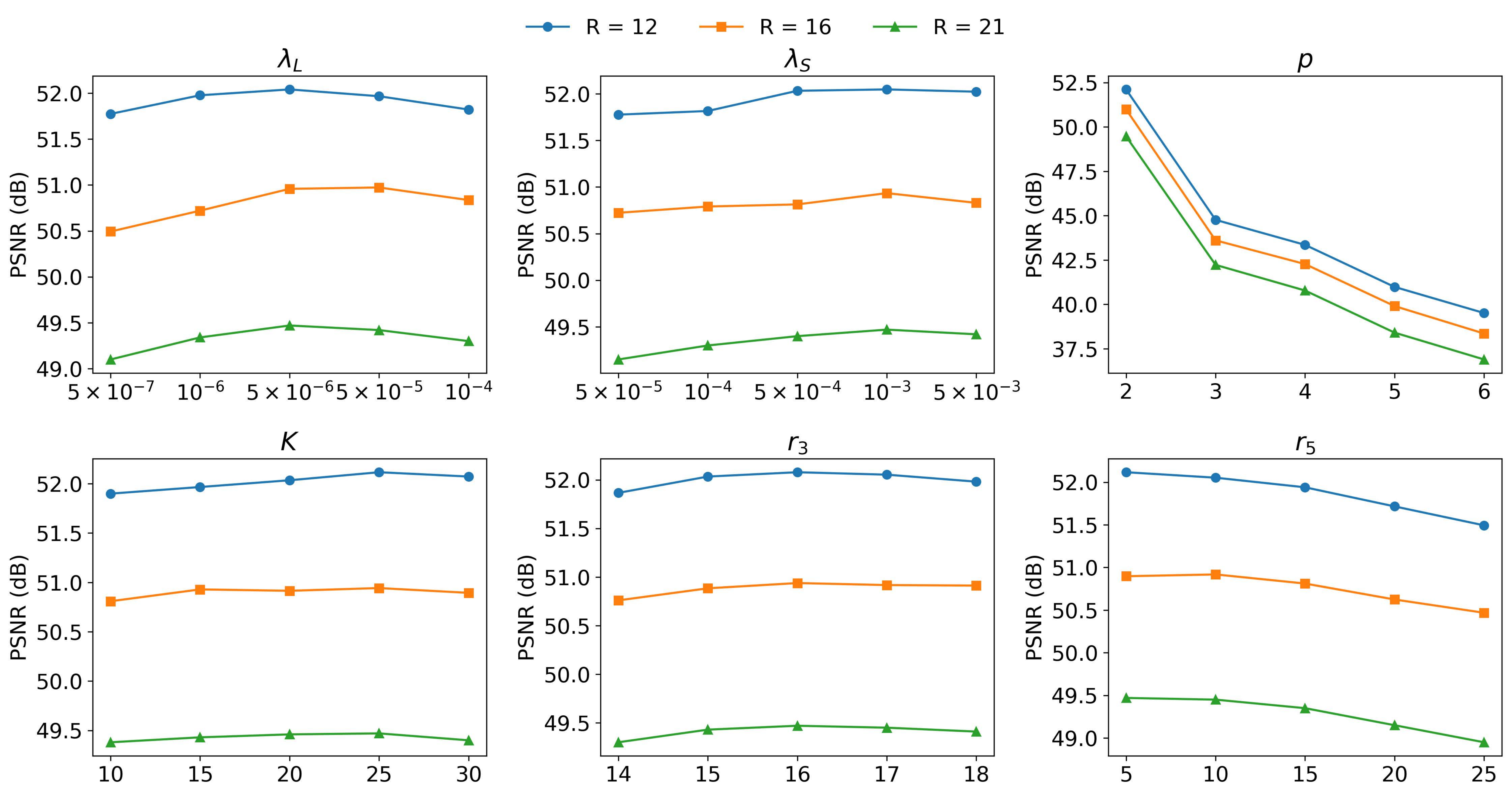}}
    \caption{Quantitative performances of TenF-INR under acceleration factors R = 12, 16, and 21, across different values of hyperparameters 
   \textit{$\lambda_{L}$, $\lambda_{S}$, $p$, $K$, $r_{3}$, and $r_{5}$.}}
    \label{fig:ab_para}
\end{figure*}

\begin{table}
\caption{Comparison of model parameters, training and inference times, and per-step training time across all  comparison methods. 
}
\label{table 2:computational performance}
\resizebox{\columnwidth}{!}{%
\begin{tabular}{c|ccccc}
\hline \hline
Method & \ Params & Training time & Inference time &  Training time per step\\ \hline
ConvDec     & 4909k  & 5.5min    & \textbf{0.004s}  & \textbf{0.015s}      \\ 
CineJENSE  & 14251k & \textbf{0.7min} & 0.033s  & 0.034s         \\
MoDL       & 339k   & 49.5h     & 0.766s         & 0.551h             \\ 
Learned DC & 329k   & 16.1h     & 0.319s         &  0.264h            \\ 
L+S-Net    & \textbf{328k} & 40.5h     &1.634s   &  0.466h           \\
TenF-INR   & 5982k  & 3.3min    &  \textbf{0.004s} & \textbf{0.015s}     \\ \hline \hline

\end{tabular}
}
\end{table}

\begin{figure}[!htbp]
\centering
\includegraphics[width=\columnwidth]{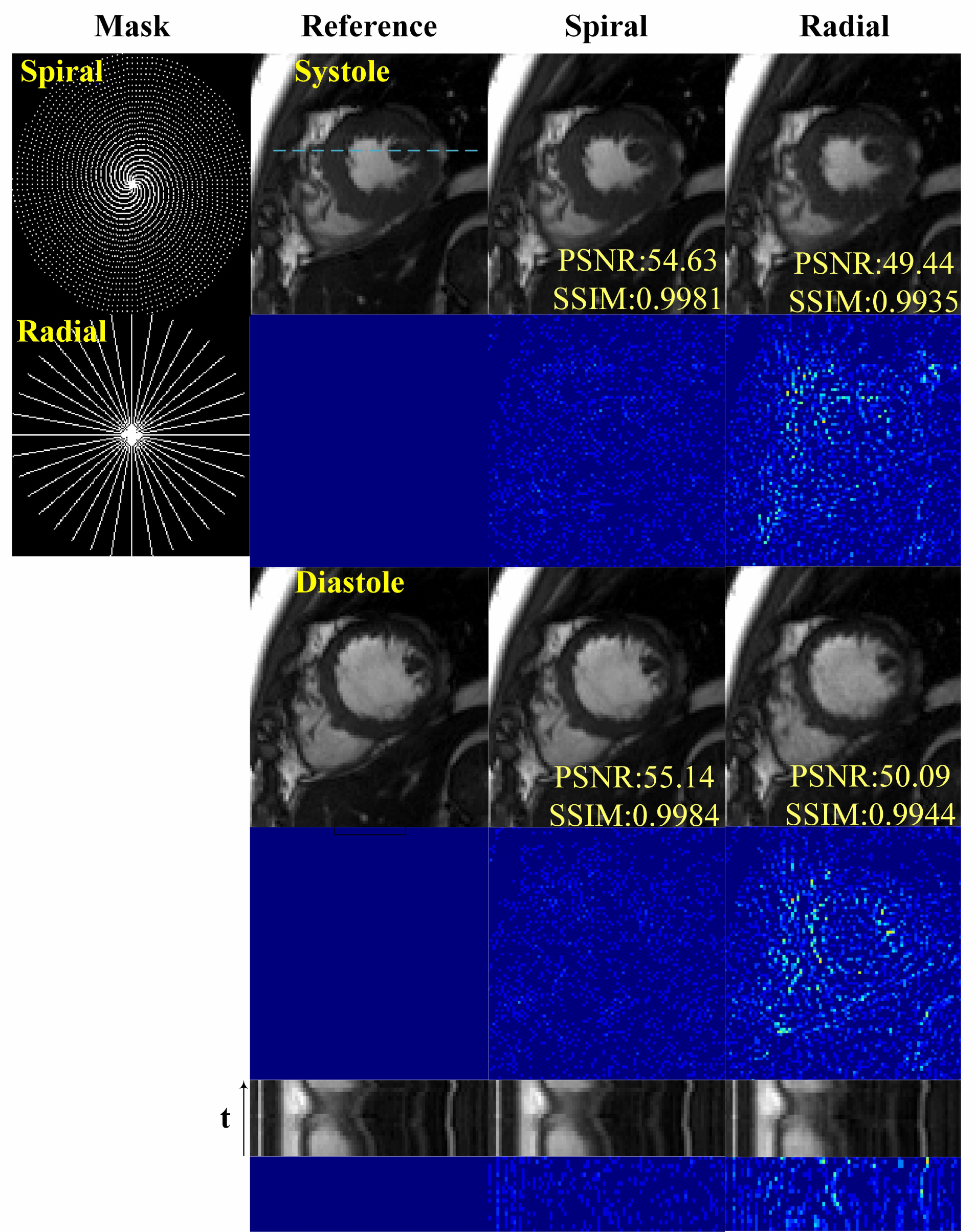}
\caption{Reconstruction results of the proposed TenF-INR method using pseudo-spiral and pseudo-radial masks at an acceleration factor of R = 12. The first four rows show the reconstructed images at diastole
and systole frames, along with the corresponding PSNR/SSIM values and error maps. The bottom two rows present the y-t dynamic image and its error map, extracted along the blue dashed line.}
\label{fig:radia_spiral}
\end{figure}

As detailed in Subsection~\ref{subsec:Implementation Details}, setting $r_1=r_2=p$ yielded optimal reconstruction, potentially because the image patches lack a low-rank structure in the spatial dimensions. Additionally, to mitigate the high computational demand of using distinct $f_{\theta_i}$ for each non-local tensor, shared parameters for $f_{\theta_i}$ were employed to model all non-local tensors, which, however, may result in a rank higher than the intrinsic one.
\par To evaluate the sensitivity of the TenF-INR method to variations in hyperparameters, we conducted controlled experiments using a retrospective dataset across all acceleration factors of R = 12, 16, and 21. For each parameter-$\lambda_{S}$, $\lambda_{L}$, $K$, $p$, $r_3$, and $r_5$-we tested 5 uniformly spaced values within predefined ranges, while maintaining the remaining parameters at their baseline values: $\lambda_{S} = 1 \times 10^{-3}$, $\lambda_{L} = 5 \times 10^{-6}$, $K = 20$, $p = 2$, $r_3 = 16$, and $r_5 = 5$. The variations in PSNR as a function of these parameter settings are shown in Fig.~\ref{fig:ab_para}. As $\lambda_{L}$, $K$, $\lambda_{S}$, and $K$ increase, PSNR exhibits a rising trend that gradually stabilizes, indicating an optimal range for enhancing reconstruction quality. 
In contrast, for $p$ and $r_5$, PSNR declines as their values increase, suggesting a trade-off where higher values may introduce artifacts or over-regularization.

\subsection{Robustness Across Diverse Undersampling Patterns }
 The proposed TenF-INR method demonstrates flexibility in reconstructing images from data acquired with various undersampling strategies. To assess its reconstruction performance, we evaluated the method using two distinct undersampling masks: pseudo-radial and pseudo-spiral at an acceleration factor of R = 12 on a retrospective cine dataset~\cite{XD-GRASP,pruessmann2001advances}. The reconstructed images at the systolic and diastolic phases are shown in Fig.~\ref{fig:radia_spiral}. It can be observed that the proposed method consistently delivers high-quality reconstructions, underscoring its robustness and adaptability across different undersampling patterns. 
 
\subsection{Computational Performance Analysis}
\par The proposed TenF-INR method offers improved computational efficiency over conventional INR-based methods. Directly modeling a non-local tensor \( \mathcal{X}_l \in \mathbb{R}^{n_1 \times n_2 \times n_3 \times n_4 } \) with conventional INRs costs \( \mathcal{O}(h^2 b n_1 n_2 n_3 n_4) \), where $h$ and $b$ are the MLP width and depth. In contrast, TenF-INR utilizes INRs to model a compact representation through tensor decomposition, representing the tensor with a parameterized core tensor and associated factor functions modeled by INRs. This reduces the cost to $\mathcal{O}(h b\hat{r} (n_1 + n_2 + n_3 + n_4) + \hat{r}n_1 n_2 n_3 n_4)$, where $\hat{r}$ is the mode-wise max rank, leading to more efficient reconstruction. This computational advantage is also evident in the reduced number of parameters in the reconstruction models. Table~\ref{table 2:computational performance} summarizes the number of parameters, training, inference, and per step training times for all comparison methods evaluated in this study. Supervised learning-based methods such as MoDL, Learned DC, and L+S-Net, which rely on conventional convolutional neural networks, have fewer parameters but demand substantially longer training times. Among INR-based methods, CineJSENSE requires approximately twice as many parameters as TenF-INR, but converges more quickly due to the use of hash encoding. In contrast, our method employs a simple MLP without any parametric encoding, in order to isolate and evaluate the performance of the INR itself, which leads to a longer training time. Several strategies can potentially accelerate convergence and further improve the performance of  TenF-INR, such as network weight initialization methods based on meta-learning \cite{meta} or cross-data transfer learning \cite{vyas2024learning}, and parametric encoding, which will be investigated in our future work.

\section{Conclusion}
This study proposes an unsupervised patch-based  reconstruction method for dynamic MRI using learnable tensor function with implicit neural representation which effectively exploits the high correlations of images in the spatiotemporal domain.
Experimental results in both retrospective and prospective imaging scenarios demonstrate that the TenF-INR method improves reconstruction both qualitatively and quantitatively, achieving superior performance in artifact suppression and image detail preservation compared to state-of-the-art methods.

\bibliographystyle{IEEEtran}  
\bibliography{refs}           

\end{document}